\documentclass[%
 aip,
rsi,
 amsmath,amssymb,
 reprint,%
]{revtex4-1}

\usepackage{graphicx}
\usepackage{dcolumn}
\usepackage{bm}

\usepackage[utf8]{inputenc}
\usepackage[T1]{fontenc}
\usepackage{mathptmx}
\usepackage{etoolbox}

\makeatletter
\def\@email#1#2{%
 \endgroup
 \patchcmd{\titleblock@produce}
  {\frontmatter@RRAPformat}
  {\frontmatter@RRAPformat{\produce@RRAP{*#1\href{mailto:#2}{#2}}}\frontmatter@RRAPformat}
  {}{}
}%
\makeatother
\begin{document}

\preprint{AIP/123-QED}
\title[]{Precision positioning in free-space optical communication systems via PID control tuned by RL}
\author{K. Prikhodko}
\author{S. Kuznetsov}
\affiliation{JSC Mostcom, 390000 Russia, Ryazan
}%

\author{S. Vorobey}%
\author{A. Katanskiy}%
\author{V. Balakirev}%
\author{A. Reutov}
 \email{ar@rootml.com}
 \altaffiliation[Also at ]{Moscow Center for Advanced Studies, Moscow, Russia}
\affiliation{LLC Science Trends, 119331 Moscow, Russia
}%

\date{\today}

\begin{abstract}
        Accurate positioning of optical components is essential for maintaining beam alignment in free-space optical (FSO) communication systems. This work investigates reinforcement-learning-assisted tuning of cascaded position and velocity PID controllers for an optical deflector that moves the end of an optical fiber in the focal plane of an optical system. A Deep Deterministic Policy Gradient (DDPG) agent adjusts six PID coefficients through interaction with a physical experimental stand. The stand supports target-coordinate updates of up to $12$ kHz, while the agent and the controlled device are located approximately $200$ km apart and exchange data over UDP. After $5000$ training sessions, two fixed coefficient sets are selected and compared with a manually tuned baseline. For a pseudo-random target trajectory, the best RL-tuned set reduces the range of the radial positioning error from $119$ to $82$, corresponding to a $31\%$ reduction, and decreases its standard deviation from $15$ to $12$. For a constant zero target, the RL-tuned sets do not improve the radial error range. The results demonstrate the potential of DDPG for experimental PID tuning in dynamic positioning tasks and indicate the need for multi-regime optimization to achieve consistent performance under different operating conditions.
\end{abstract}

\maketitle


\section{Introduction}

Free-space optical (FSO) communication systems require accurate alignment of transmitting and receiving optical components. Small beam displacements in the focal plane may reduce coupling into the receiving fiber and degrade the reliability of the communication link. Precision positioning devices are therefore an important part of FSO terminals, particularly when mechanical vibrations, platform motion, or other time-dependent disturbances must be compensated. In this work, positioning is performed by an optical deflector (OD) that moves the end of an optical fiber in the focal plane of the receiving optical system \cite{Boev2021}. The device accepts target-coordinate sequences at frequencies of up to $12$ kHz and must reproduce them with a small tracking error.

Proportional-integral-derivative (PID) control remains widely used in industrial motion and process-control applications because of its simple structure, low computational cost and interpretability \cite{Borase2021PIDreview, Johnson2005PID}. However, control quality depends strongly on the selected proportional, integral, and derivative coefficients. Manual procedures such as the Ziegler-Nichols method \cite{Ziegler1942} provide useful initial settings, but additional experimental adjustment may be required in systems affected by nonlinearities, communication delays, measurement noise, or actuator constraints. This motivates automated tuning procedures that can learn directly from data collected on the controlled device.

Reinforcement learning (RL) provides a framework for learning control decisions through interaction with an environment and has been applied to a broad range of control problems, including autonomous driving \cite{Driving2017, Driving2021} and energy systems \cite{Wei, Energy}. Nevertheless, replacing a classical controller with an end-to-end neural policy is not always justified, since conventional methods remain competitive and often provide more predictable behavior \cite{Bohn2019}. A practical alternative is a hybrid approach in which RL does not replace the PID controller but assists in selecting or adapting its coefficients.

The combined DDPG-PID approach was designed and applied to the inverse pendulum task \cite{DRPID}: in this article authors provided comparative analysis of only-PID, only-DDPG and DDPG-PID control schemes. Further, authors was investigated \cite{SACaircraft2023} morphing aircaft strategy based on the reinforcement learning and there was used on of the actor-critic model similar to DDPG, the Soft Actor Critic (SAC) model. Authors verified safety and feasibility of deployment and compared their scheme with genetic algorithm and PPO \cite{PPO} strategy. The new approach with episodic tuning framework and self-preserving method was introduced in the article \cite{Self-preserving}. Here authors provided combination of Deterministic Policy Gradient method with PID control strategy for optimization of a modeled system described by second-order plus dead-time process. Another hybrid DDPG-PID approach was applied \cite{MIMOPID} for simulated robotic platform and a real-time mobile robot. These results suggest that RL can act as a data-driven tuning layer around a conventional feedback controller. At the same time, the effectiveness of such methods depends on the physical plant, the reward definition, safety constraints and the difference between simulated and real operating conditions.

The present study examines RL-assisted PID tuning for a high-frequency optical positioning device on a physical experimental stand. The OD is controlled by cascaded position and velocity PID loops, and the RL action contains six coefficients, $\{P_p,I_p,D_p,P_v,I_v,D_v\}$, applied to the controllers of both coordinate axes. We use the Deep Deterministic Policy Gradient (DDPG) algorithm because the tuning problem has a continuous action space and because this class of actor--critic methods has previously been used for PID parameter optimization \cite{DDPG1,DDPG2,DRPID,Self-preserving,MIMOPID}. The agent interacts directly with the physical OD rather than with a numerical plant model. The environment is located in Ryazan, while the agent runs in Moscow, approximately $200$ km away; target coordinates, measured positions and controller coefficients are exchanged over UDP communication protocol. This distributed configuration introduces communication delays and occasional connection failures in addition to the dynamics of the OD.

The goal is to use RL as an experimental search procedure for improved fixed PID coefficients, rather than to replace the existing controller with a neural policy. The DDPG agent is trained for $5000$ communication sessions, after which two candidate coefficient sets are selected from the recorded history. They are compared with a manually tuned baseline while holding the fiber end at the constant target $\{0,0\}$ and while tracking a pseudo-random sequence matched to the frequency response of the OD. For the dynamic trajectory the best RL-tuned set reduces the range of the radial positioning error by $31\%$, from $119$ to $82$, and its standard deviation from $15$ to $12$. For the constant target, the radial range changes from $13$ to $14$, so no improvement is observed. The results therefore demonstrate a benefit for the tested dynamic regime, but not a universal improvement across operating conditions.

This work is structured as follows:
\begin{itemize}
      \item A description of the reinforcement learning model is introduced in Section \ref{sec:ml}.
      \item Section \ref{sec:stand} provides information on the experimental setup and the tuning of PID controllers.
      \item We provide test results of precise positioning with found and manually obtained PID coefficients in Section \ref{sec:tests}.
      \item Finally, in Section \ref{sec:conc} we briefly summarize our paper and conclude with a discussion of future research.
\end{itemize}

\section{Reinforcement learning model}\label{sec:ml}
Reinforcement learning is formulated as a sequential interaction between an agent and an environment. At each decision step $t$, the agent operates in some environment and this environment is described by state $s_t$. The agent strives to perform an action $a_t$ that is beneficial to him, receiving feedback on the performed action in the form of a numerical reward $r_t$. Eventually, the executed action and the internal dynamics of the environment change the state to a new one $s_{t`}$. By repeatedly interacting with the environment, the agent learns a control policy that maximizes the expected cumulative reward rather than the reward obtained at an individual step.

The agent's decision-making process can described as a Markov (or semi-Markov \cite{Sutton1999Semi-Markov}) process. We choose semi-Markov approach via various instabilities in communication channel between two distributed points (the environment and the agent) and high-frequency of operating mode of OD, which is a main part of environment experimental setup. Control policy is formally defined a value that indicates what action $a_t$ the agent needs to perform under a given state $s_t$ of environment. To build a policy, Q-value adjusted by the Bellman equation can be used:
\begin{equation}
    Q(a_t, s_t) = r_t + \gamma\, \underset{a_{t'}}{\rm max}[Q(a_{t'},s_{t'})],
    \label{eq:Bellman}
\end{equation}
where $\gamma$ is the discount factor (a parameter of the reinforcement learning model, $\gamma \in (0,1)$). This recurrent formula allows to recalculate the Q-value in the presence of new data from the agent's experience $\{s_t,\,a_t,\,r_t,\,s_{t'}\}$. The policy can be constructed in different ways: for example, an action $a_t$ is selected in which the Q-value $Q(a_t, s_t)$ reaches a maximum at a given state $s_t$. The Bellman equation ensures that the policy thus constructed converges to the optimal one, i.e. accumulated reward $\sum \gamma^{t'} r_{t'}$ is maximized for each executed action $a_t$ in state $s_t$ during decision process.

For our problem -- the fine-tuning of PID controllers of OD -- we choose the algorithm of deep deterministic gradient policy (DDPG) \cite{DDPG1,DDPG2}. This algorithm provides continuous control, which allows using a wide range of practical and classical control problems, in particular, the task of PID coefficient manipulation \cite{DRPID,Self-preserving, MIMOPID}. There are other types of RL models suitable for continuous control (PPO \cite{PPO}, TRPO \cite{TRPO}, SAC \cite{SAC}). We choose DDPG because of simpler structure, bigger history of industrial usage and competitive benchmark results \cite{NIAN2020, Henderson2018}. The implementation of the DDPG model largely repeats the approaches already incorporated in the DQN (Deep Q-learning Network) algorithm, which is tested in wide range of discrete control tasks \cite{mnih2015humanlevel}. In the software development level DDPG differs from DQN implementations by several conceptually simple changes: there is realization of two neural networks, the "actor", a policy neural network that calculates the optimal action $a_t$ for a given state $s_t$, and the "critic", a Q-value estimation neural network, that specifies a metric for the “actor” training.

For the PID coefficient control problem we define state $s_t$, action $a_t$, and reward $r_t$:
\begin{itemize}
     \item \textbf{State}. This is a set of values of the current position ($x_{{\rm pos}, i}$ and $y_{{\rm pos},i}$) and the target position ($x_{{\rm target}, i}$ and $ y_{{\rm target},i}$) of OD. One state is formed from a sequence of 362 points $\{x_{{\rm pos}, i},\,y_{{\rm pos}, i},\,x_{{\rm target}, i}, \,y_{{\rm target}, i}\}$. This number of points is chose due to limits of packets size of the chosen in our experiment communication protocol UDP.
     \item \textbf{Action}. The action $a_t$ is defined as the change in the PID controller coefficients along the coordinates $\{P_p,\,I_p,\,D_p\}$ and the PID controller coefficients along the speeds $\{P_v,\, I_v,\,D_v\}$ . The same set of coefficients is transmitted to a pair of PID controllers along the $Y$ axis as along the $X$ axis - in total, the action is described by six numbers $a_t = \{P_p,\,I_p,\,D_p \,P_v\,I_v,\, D_v\}$. It is important to note, that whole control scheme have two PID controlers: two along the coordinates (axis $X$ and axis $Y$ respectively) and two along the speeds (along axis $X$ and along axis $Y$ respectively). Overall we have 2-PID control units for every axis. In oue setup We imply the same sets of coefficients $a_t$ both for this 2-PID control units. 
     \item \textbf{Reward}. The reward is defined as the sum of $N=362$ distance values between the current positions and the target positions:
     \begin{equation}
         r = k_r\sum^{N}_{i=1}\sqrt{(x_{{\rm pos}, i}-x_{{\rm target}, i})^2+(y_{{\rm pos}, i}-y_{{\rm target}, i})^2},
     \end{equation}
     where $k_r$ is a normalization coefficient.
\end{itemize}

\textbf{$\epsilon$-greedy exploration}. Our model implements the $\epsilon$-greedy exploration strategy as follows: a random term of order $\epsilon \cdot a_t$ is added to each action $a_t$. During the training process the value $\epsilon$ is discounted to a near-zero value $\epsilon_f$ (this value is chosen as $\epsilon_f=0.0001$). This solution allows the agent, at the beginning of training, to explore the action space $A$ (any $a_t \in A$) and obtain the more varied states $s_{t`}$, receiving more diverse reinforcement $r_t$.

\textbf{Training process}. The ADAM optimizer \cite{kingma2017adam} are used for training. The “critic” neural network of the DDPG algorithm is trained via the Bellman formula (\ref{eq:Bellman}). For “critic” training, one generally uses data recorded in the memory of the agent in following form: reward $r_t$, states $s_t$, $s_{t'}$ and the corresponding executed actions $a_t$ and $a_{t'}$. Training the neural network “actor”
occurs on state $s_t$ data using the metric specified by the “critic” neural network. The training tuple $\{s_t,\,a_t,\,r_t,\,s_{t'}\}$ is collected in the process of decision making by the neural network “actor” which execute actions $a_t$ performed in states $s_t$. Our DDPG model in the choice of hyperparameters and activation layers for critic and actor networks follows the paper by T. P. Lillicrap, et al. \cite{DDPG2} and this choice was tested for various continious control tasks \cite{Henderson2018} and was implemented for RL-PID control frameworks \cite{Self-preserving,MIMOPID}. The actor and critic  networks are designed as multilayer perceptrons with two hidden layers of sizes $80$ for first hidden layer and $60$ for second hidden layer. The parameter of the $\epsilon$-greedy exploration strategy is discounted from $\epsilon_i = 0.3$ to $\epsilon_f=0.001$ with decrement $1/1000$ during every transition between states $s_t \rightarrow s_{t'}$.


\section{Experimental setup and PID tuning}\label{sec:stand}
The training process and testing of the system are explained in this section. The system consist of experimental stand located in Ryazan (Russia, 200 km from Moscow) and the agent in Moscow connected with the stand.
\subsection{Stand}
    
\begin{figure}[ht!]\centering
	\includegraphics[width=0.45\textwidth]{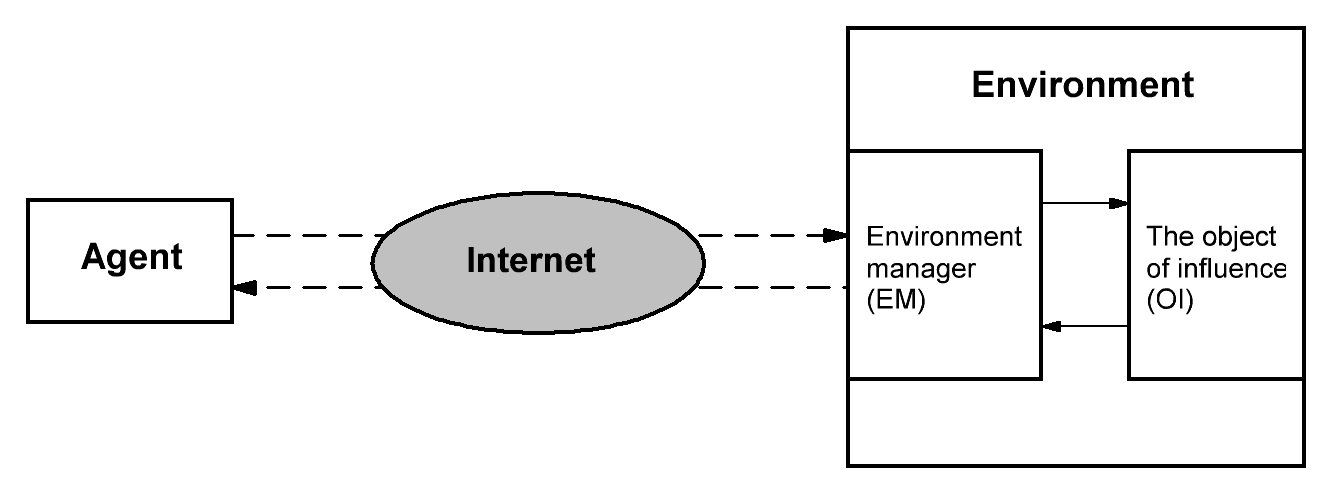}
	\caption{Scheme of the experimental stand for optimization of PID controllers of OD.}
	\label{fig:equip}
\end{figure}

The structure of the experimental stand for optimizing the coefficients of PID OD consisting of an Agent, an Environment Manager (EM) and an Object of Influence (OI) is presented in Fig. \ref{fig:equip}.

The presented stand are worked in accordance with the following algorithm:
\begin{enumerate}
     \item Environment Manager (EM) is permanently in listening on the specified port through which communication with the Agent is carried out via the Internet.
     \item The Agent network module sends a data packet consisting of a control segment $\{P_p,\,I_p,\,D_p \,P_v\,I_v,\, D_v\}$ and coordinate values  $(x_{{\rm target}, i},\, y_{{\rm target},i})$ to the agreed IP address/port to the EM address.
     \item The EM receives the packet, checks it for compliance with the format in accordance with the control mode byte, changes the PID coefficients of the OI automatic control system, sets the frequency for specifying coordinates and transmits a set of new $(x_{{\rm target}, i},\, y_{{\rm target},i})$ coordinates to OI.
     \item OI performs the specified movement, synchronously recording position sensor data and sends an array of measured positions $(x_{{\rm pos}, i},\, y_{{\rm pos},i})$ to the EM.
     \item EM generates a response packet to the Agent, containing the control signal $\{P_p,\,I_p,\,D_p \,P_v\,I_v,\, D_v\}$ previously received from it and an array of measured coordinates $(x_{{\rm pos}, i},\, y_{{\rm pos},i})$ and sends it to the Agent.
     \item The agent receives the response packet and goes into the mode of processing the received information until the next contact initiation.
\end{enumerate}

The object of influence in this stand is an optical deflector. This device allows to operate with rapid movements of the received optical beam in the focal plane by precision moving the end of the receiving optical fiber \cite{Boev2021}. 

\begin{figure}[t!]\centering
	\includegraphics[width=0.45\textwidth]{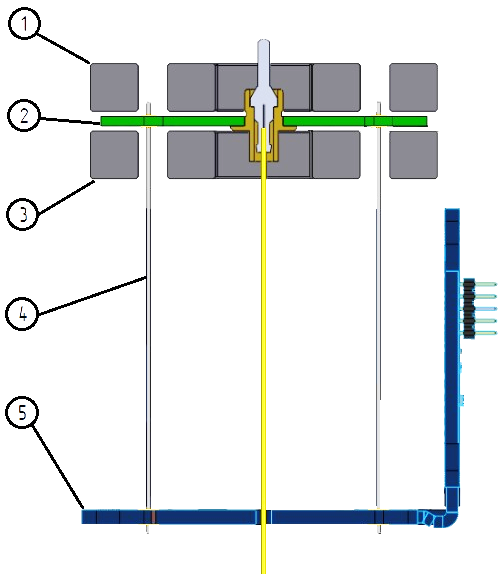}
	\caption{Scheme of an optical deflector. 1 is a permanent magnet, 2 is an drive board of optical fiber end, 3 is a permanent magnet, 4 is a traverses, 5 is a control board.}
	\label{fig:OD}
\end{figure}

 A scheme of an optical deflector is shown In Fig. \ref{fig:OD}. The drive board of optical fiber end (OFE - 2) is placed in the magnetic field formed by permanent magnets 1 and 3, which is fixed at the base of the deflector on a fixed printed circuit board by four flexible metal rods (traverses). Through the traverses 4 electric current flows to the electromagnetic coils of the X and Y axes of the movable board 2. The interaction of the coil current of the board 2 with the magnetic field of permanent magnets 1 leads to the movement of the board 2 in the plane of the base. Flexible-rigid control 5 provides current control to the coils of the moving board 2 and also includes a position feedback system.
The deflector control system allows to feed it arrays of coordinates of a target position positions $(x_{{\rm target}, i},\, y_{{\rm target},i})$ of the OFE and read the measured positions positions $(x_{{\rm pos}, i},\, y_{{\rm pos},i})$ of the OFE. The size of thus arrays is up to 65535 points with a coordinate setting frequency of up to 12 kHz.

The agent’s interaction with the environment occurs during request-response sessions through the exchange of data in UDP packets over the Internet. The interaction is initiated by the agent and the environment's port in a listening state and responds to the agent's requests. The agent sends a packet of 1498 bytes to the agreed IP address and port. The size of the UDP packet was determined in accordance with the maximum transmission unit (MTU) to achieve more effective interaction. The protocol is symmetric: the agent sends 362 target pairs to the environment and expects 362 pairs of position values in response. The differences between targets and positions are errors, which are used to determine the accuracy of the control system. The UDP packet structure is shown in the Table \ref{tab:UDP}.

\begin{table}[ht]
\begin{center}
\begin{tabular}{ |c|c|l|l| } 
    \hline
     Byte address& Length & Description & Possible values \\  
     \hline
     0& 1& Control Byte & $0$ -- baseline settings,\\ 
     &&&$1$ -- position PID control,\\
     &&&$2$ -- velocity PID control,\\
     &&&$3$ -- position and velocity\\
     &&& PID control. \\
     \hline
     1 & 48 & 12th PID & Float number\\
     && coefficients &\\
     \hline
     49&1&Frequency& Integer\\
     \hline
     50&1448&$x_{{\rm pos}, i},\, y_{{\rm pos},i}$&Integers\\
    \hline
\end{tabular}
\end{center}
  \caption{The structure of UDP packet. Control byte represents four control modes (no control from RL agent, position PID coefficients change,  velocity PID coefficients change, both position PID coefficients and velocity PID coefficients change). Frequency are chosen rounded to the hundred. $x_{{\rm pos}, i},\, y_{{\rm pos},i}$ represent 362 pairs of position (or target) values.}
\label{tab:UDP}
\end{table}

\subsection{Training on the stand and description of the tests}
During training the agent transmits to the stand an action values $a_t = \{P_p,\,I_p, \,D_p,\,P_v,\,I_v,\,D_v\}$, i.e. new PID coefficients, each communication session. Also from the agent side, we transmit $362$ pairs of target values $x_{{\rm target}, i },\,y_{{\rm target}, i}\}_{i=1}^{N=362}$.
In response, the environment transmits $362$ pairs of position values $\{x_{{\rm pos}, i},\,y_{{\rm pos}, i}\}_{i=1}^{N=362}$. Eventually, the agent gets a new state $s_{t`} = \{x_{{\rm pos}, i},\,y_{{\rm pos}, i},\,x_{{\rm target}, i },\,y_{{\rm target}, i}\}_{i=1}^{N=362}$, i.e. $1448$ target and current position values on both axes for $362$ steps. The values for the target position are also specified in the form:
\begin{equation}
     y_{\rm target}(t_i) = U(t_i) + P(t_i),
\end{equation}
where $U(x)$ is a step periodic function, $P(x)$ is a pseudo-random sequence (PRS)

corresponding to the frequency response of the OD. For this PRS a random vector was generated in polar coordinates (its length and angle, both parameters had a uniform distribution). The resulting vector was decomposed into x and y coordinates; along each axis the coordinate entered a low-frequency filter, the frequency response of which was identical to the frequency response of the OD. After the filter, the resulting array of coordinates was scaled so that the speed did not exceed the maximum for which the deflector was designed. If there was no excess, then the array was not scaled.

\begin{figure}[t!]\centering
	\includegraphics[width=0.45\textwidth]{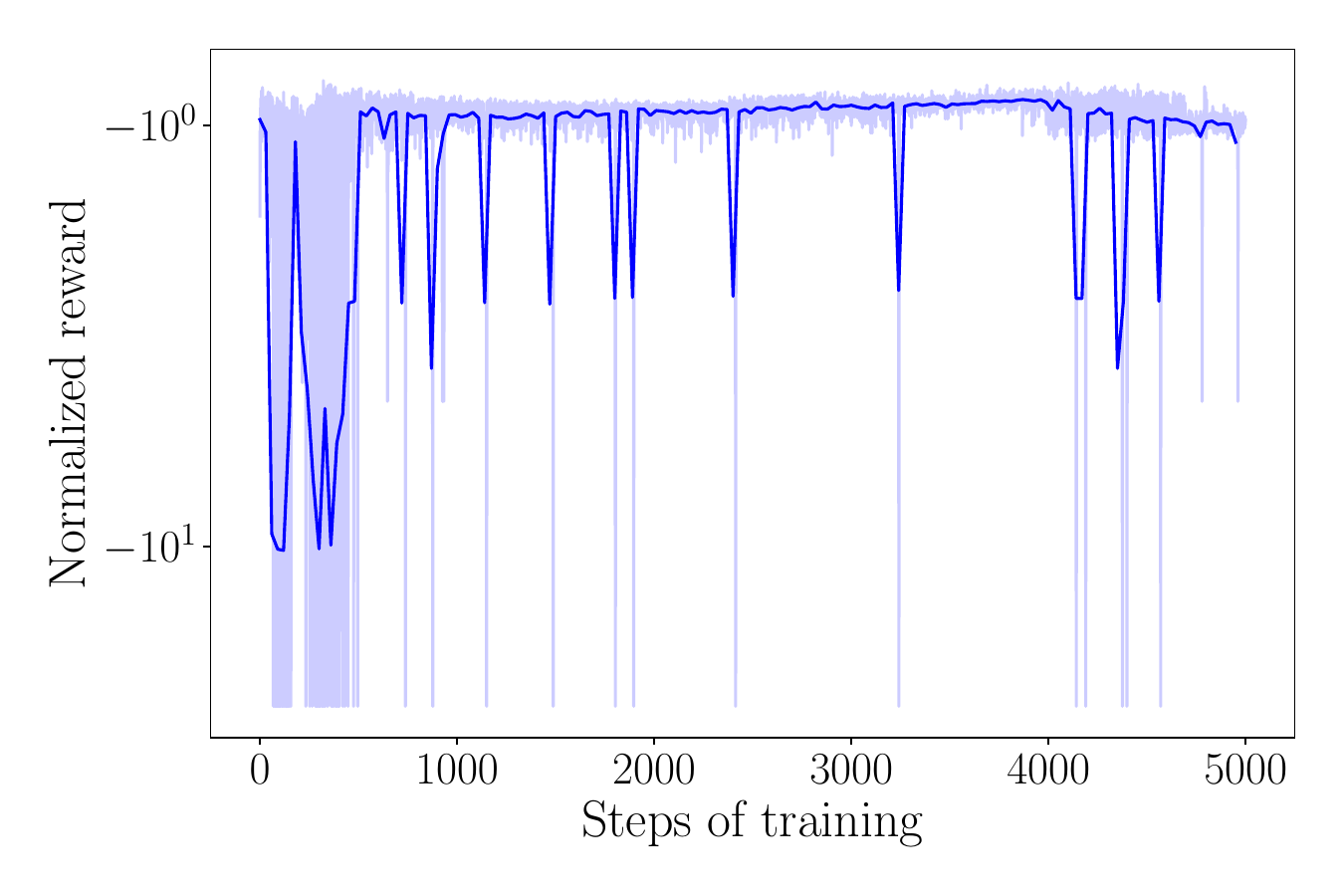}
	\caption{Reward $r_t$ for $5000$ communication sessions. Reward values are presented on a logarithmic scale. The pale blue line represents the raw reward data, the bright blue line represents the  reward averaged over 30 values. The spikes of reward value mostly represents the self-preserving penalty, which occurs if error exceeds reward value obtained for baseline PID control. Another reason of such behaviour is connection failures: the information about environment state are not received to agent and penalty equal self-preserving penalty applied as whole reward for agent.}
	\label{fig:reward_function}
\end{figure}

In total, the agent is trained on $5000$ sessions before the lasting increase of the reward function. The $\epsilon$-greedy exploration strategy stops after the first $1000$ communication sessions. The reward metrics of the training is shown in Fig. \ref{fig:reward_function}.

The entire history of the training process is saved. We determine the optimal coefficients from the received training data. The optimality of the action is estimated by the maximum values of the reward $r_t$. For PID coefficients, two optimal sets $a_{\rm opt, 1}$ and $a_{\rm opt, 2}$ is selected (see Table \ref{tab:a_opt}).
\begin{table}[ht]
\begin{center}
\begin{tabular}{ |c|c|c|c|c|c|c| } 
    \hline
     &$P_p$ & $I_p$ & $D_p$ &$P_v$&$I_v$&$D_v$ \\  
    \hline
    $a_{\rm opt, 1}$ & 4.602 &0.08649&12.29& 3.238&0.0004861&14.34 \\ 
    \hline
     $a_{\rm opt, 2}$ &4.676&0.08508&12.95&3.119& 0.0004785&15.32 \\ 
    \hline
\end{tabular}
\end{center}
  \caption{Found sets of coefficients $a_{\rm opt, 1}$ and $a_{\rm opt, 2}$ for PID controllers $\{P_p,\, I_p,\, D_p\}$ along coordinates and PID controllers along speed $\{P_v,\, I_v,\, D_v\}$.}
\label{tab:a_opt}
\end{table}

We test our experimental stand with the baseline coefficients $a_{\rm baseline}$ and the estimated optimal coefficients $a_{\rm opt, 1}$ and $a_{\rm opt, 2}$.
PRS corresponding to the frequency response of the OD is selected as sequence of target values, and tests is also carried out for constant target values $\{0.0\}$.
For all measures the following numerical values are determined:
\begin{itemize}
     \item value $\Delta X_i = x_{{\rm pos}, i}-x_{{\rm target}, i}$;
     \item value $\Delta Y_i = y_{{\rm pos}, i}-y_{{\rm target}, i}$;
     \item value $\Delta R_i = \sqrt{(x_{{\rm pos}, i}-x_{{\rm target}, i})^2+(x_{{\rm pos}, i}-x_ {{\rm target}, i})^2}$.
\end{itemize}
For each test numerical characteristics (standard deviation $\sigma$ and range of values $D$) are calculated for these experimentally measured value $\Delta X_i, \, \Delta Y_i, \, \Delta R_i$.

\section{Test results}\label{sec:tests}
The tests provide a comparison of the OD operation with the baseline set of coefficients and with those obtained from the training results $a_{\rm opt, 1}$ and $a_{\rm opt, 2}$. The values of the manually tuned baseline coefficients $a_{\rm baseline}$ are presented in the Table \ref{tab:a_baseline}.

\begin{table}[ht]
\begin{center}
\begin{tabular}{ |c|c|c|c|c|c|c| } 
    \hline
     &$P_p$ & $I_p$ & $D_p$ &$P_v$&$I_v$&$D_v$ \\  
    \hline
    $a_{\rm baseline}$ & 4 &0.07&20& 3&0.0004&20 \\ 
    \hline
\end{tabular}
\end{center}
  \caption{Baseline set of coefficients $a_{\rm baseline}$ for PID controllers $\{P_p,\, I_p,\, D_p\}$ along coordinates and PID controllers along speed $\{P_v,\, I_v,\, D_v\}$.}
\label{tab:a_baseline}
\end{table}
We analyze the deviations of the measured position from the specified one when the OFE is held at the point $(0,0)$, as well as when a PRS is applied to it. In the test, the PRS  is the sequence of coordinates with a size in 65535 points, the frequency of target value change is set at level 12 kHz.
During testing we measure the values of standard deviation $\sigma$ and range $D$ for each coordinate axes $X$, $Y$ and for distance $R$ and this averaged measured results are given in the Table \ref{tab:results}.
\begin{table}
\begin{center}
\begin{tabular}{ |c|c|c|c|c|c| } 
    \hline
     Test & PID & ${\sigma}_X/{D}_X$ & ${\sigma}_Y/{D}_Y$  &${\sigma}_R/{D}_R$  & Variance of ${D}_R$\\  
    \hline
    Target $\{0,0\}$ &$a_{\rm baseline}$ &2.3 / 23 &2.4 / 23 &1.8 / 13 &1.1\\ 
    \cline{2-6}
     &$a_{\rm opt, 1}$ &3.3 / 27	&2.3 / 29 &2.2 / 14	&1\\ 
    \cline{2-6}
    &$a_{\rm opt, 2}$ &2.9 / 25	&2.6 / 25	&2.1 / 14	&5.7\\ 
    \hline
    Target $P(x)$ &$a_{\rm baseline}$& 21 / 212	&22 / 212	&15 / 119	&88\\ 
    \cline{2-6}
     &$a_{\rm opt, 1}$ &17 / 153	&18 / 159	&12 / 82	&41\\ 
    \cline{2-6}
    &$a_{\rm opt, 2}$ &18 / 172	&19 / 175	&12 / 103	&198\\ 
    \hline

\end{tabular}
\end{center}
  \caption{Table of standard deviation $\sigma$ and range of values $D$ for sets $a_{\rm baseline}$, $a_{\rm opt, 1}$ and $a_{\rm opt, 2}$ and target selection by PRS $P(x)$ or stable zero value $\{0,0\}$.}
\label{tab:results}
\end{table}
As shown in the Table \ref{tab:results}, for the case of zero-position target values, the new sets of coefficients increase the ranges $D_{R}$ by approximately 7.7\% 
compared with the basic set. 


For PRS, $D_{R}$ decreases by 31\% and 13.4\% for $a_{\rm opt, 1}$ and $a_{\rm opt, 2}$, respectively.
The proposed sets of coefficients differ from the baseline in terms of variance of range $D_R$: the stand operation with $a_{\rm opt, 1}$ and PRS has a reduction of this metric by half, $a_{\rm opt, 2}$ increases it by 2.25 times.
Thus, for this measurements the $a_{\rm opt, 1}$ set can be considered the more optimal than baseline set $a_{\rm baseline}$.

\section{Summary}\label{sec:conc}

In this work, we investigated the application of reinforcement learning to the experimental tuning of cascaded position and velocity PID controllers used in a precision optical positioning system. A DDPG agent was employed to search in a continuous six-dimensional space of PID coefficients through direct interaction with a physical experimental stand. In contrast to approaches based exclusively on numerical simulation, the proposed training procedure included the dynamics of the real controlled device, measurement noise, communication delays and occasional failures of the distributed communication channel. The reinforcement learning agent was used as a parameter-tuning mechanism rather than as a direct replacement for the conventional feedback controller.

After $5000$ communication sessions, two candidate sets of PID coefficients, $a_{\rm opt,1}$ and $a_{\rm opt,2}$, were selected from the recorded training history and compared with the manually tuned baseline set $a_{\rm baseline}$. The comparison was performed for two operating regimes: maintaining a constant target position and tracking a pseudo-random target sequence corresponding to the frequency response of the controlled device. For the pseudo-random sequence, the coefficient set $a_{\rm opt,1}$ reduced the range of the radial positioning error $D_R$ from $119$ to $82$, which corresponds to a reduction of approximately $31\%$. The standard deviation of the radial error was also reduced from $15$ to $12$. The second candidate set, $a_{\rm opt,2}$, produced a smaller reduction of the radial error range, from $119$ to $103$.

For the constant target position, however, neither of the RL-tuned coefficient sets improved the radial error range relative to the baseline. This result indicates that the coefficients obtained by optimizing performance for a dynamic target trajectory are not necessarily optimal for a stationary operating regime. Therefore, the presented experiments demonstrate the feasibility of DDPG-assisted PID tuning for the considered dynamic positioning task, but they do not establish a universal improvement under all operating conditions. The results also emphasize the importance of selecting a training trajectory and reward function that adequately represent the expected operating regimes of the control system.

Further research should consider PID tuning from a wider range of initial coefficient values, including training ``from scratch'' without manually selected parameters close to the baseline solution. The reward function may also be extended to account simultaneously for tracking accuracy, stationary-position accuracy, oscillations, control effort and transitions between different operating regimes. Additional experiments are required to evaluate the repeatability of the obtained coefficient sets under variations in external disturbances, mechanical stress, communication latency, and hardware characteristics. A further direction is the development of constrained adaptive tuning, in which PID coefficients are updated during operation while remaining within predefined safe ranges. Such an approach may allow the controller to compensate for gradual changes in the system dynamics while preserving the interpretability and reliability of the underlying PID control structure.

\hfill


\section*{Data Availability Statement}
The data that support the findings of this study are available from the corresponding author upon reasonable request.


\begin{thebibliography}{23}%
\makeatletter
\providecommand \@ifxundefined [1]{%
 \@ifx{#1\undefined}
}%
\providecommand \@ifnum [1]{%
 \ifnum #1\expandafter \@firstoftwo
 \else \expandafter \@secondoftwo
 \fi
}%
\providecommand \@ifx [1]{%
 \ifx #1\expandafter \@firstoftwo
 \else \expandafter \@secondoftwo
 \fi
}%
\providecommand \natexlab [1]{#1}%
\providecommand \enquote  [1]{``#1''}%
\providecommand \bibnamefont  [1]{#1}%
\providecommand \bibfnamefont [1]{#1}%
\providecommand \citenamefont [1]{#1}%
\providecommand \href@noop [0]{\@secondoftwo}%
\providecommand \href [0]{\begingroup \@sanitize@url \@href}%
\providecommand \@href[1]{\@@startlink{#1}\@@href}%
\providecommand \@@href[1]{\endgroup#1\@@endlink}%
\providecommand \@sanitize@url [0]{\catcode `\\12\catcode `\$12\catcode `\&12\catcode `\#12\catcode `\^12\catcode `\_12\catcode `\%12\relax}%
\providecommand \@@startlink[1]{}%
\providecommand \@@endlink[0]{}%
\providecommand \url  [0]{\begingroup\@sanitize@url \@url }%
\providecommand \@url [1]{\endgroup\@href {#1}{\urlprefix }}%
\providecommand \urlprefix  [0]{URL }%
\providecommand \Eprint [0]{\href }%
\providecommand \doibase [0]{http://dx.doi.org/}%
\providecommand \selectlanguage [0]{\@gobble}%
\providecommand \bibinfo  [0]{\@secondoftwo}%
\providecommand \bibfield  [0]{\@secondoftwo}%
\providecommand \translation [1]{[#1]}%
\providecommand \BibitemOpen [0]{}%
\providecommand \bibitemStop [0]{}%
\providecommand \bibitemNoStop [0]{.\EOS\space}%
\providecommand \EOS [0]{\spacefactor3000\relax}%
\providecommand \BibitemShut  [1]{\csname bibitem#1\endcsname}%
\let\auto@bib@innerbib\@empty
\bibitem [{\citenamefont {Boev}\ \emph {et~al.}(2021)\citenamefont {Boev}, \citenamefont {Kernosov}, \citenamefont {Kuznetsov}, \citenamefont {Parshin}, \citenamefont {Polyakov},\ and\ \citenamefont {Shirobakin}}]{Boev2021}%
  \BibitemOpen
  \bibfield  {author} {\bibinfo {author} {\bibfnamefont {A.~A.}\ \bibnamefont {Boev}}, \bibinfo {author} {\bibfnamefont {M.~Y.}\ \bibnamefont {Kernosov}}, \bibinfo {author} {\bibfnamefont {S.~N.}\ \bibnamefont {Kuznetsov}}, \bibinfo {author} {\bibfnamefont {A.~A.}\ \bibnamefont {Parshin}}, \bibinfo {author} {\bibfnamefont {S.~Y.}\ \bibnamefont {Polyakov}}, \ and\ \bibinfo {author} {\bibfnamefont {S.~E.}\ \bibnamefont {Shirobakin}},\ }in\ \href@noop {} {\emph {\bibinfo {booktitle} {Trudy XXIX Mezhdunarodnoj Konferencii «Lazerno-informacionnye tekhnologii» [Proceedings of the XXIX International Conference «Laser Information Technologies»]}}}\ (\bibinfo {year} {2021})\BibitemShut {NoStop}%
\bibitem [{\citenamefont {Borase}\ \emph {et~al.}(2021)\citenamefont {Borase}, \citenamefont {Maghade}, \citenamefont {Sondkar},\ and\ \citenamefont {Pawar}}]{Borase2021PIDreview}%
  \BibitemOpen
  \bibfield  {author} {\bibinfo {author} {\bibfnamefont {R.}~\bibnamefont {Borase}}, \bibinfo {author} {\bibfnamefont {D.}~\bibnamefont {Maghade}}, \bibinfo {author} {\bibfnamefont {S.}~\bibnamefont {Sondkar}}, \ and\ \bibinfo {author} {\bibfnamefont {S.}~\bibnamefont {Pawar}},\ }\href {\doibase 10.1007/s40435-020-00665-4} {\bibfield  {journal} {\bibinfo  {journal} {International Journal of Dynamics and Control}\ }\textbf {\bibinfo {volume} {9}} (\bibinfo {year} {2021}),\ 10.1007/s40435-020-00665-4}\BibitemShut {NoStop}%
\bibitem [{\citenamefont {Johnson}\ and\ \citenamefont {Moradi}(2005)}]{Johnson2005PID}%
  \BibitemOpen
  \bibfield  {author} {\bibinfo {author} {\bibfnamefont {M.~A.}\ \bibnamefont {Johnson}}\ and\ \bibinfo {author} {\bibfnamefont {M.~H.}\ \bibnamefont {Moradi}},\ }\href@noop {} {\emph {\bibinfo {title} {PID control}}}\ (\bibinfo  {publisher} {Springer},\ \bibinfo {year} {2005})\BibitemShut {NoStop}%
\bibitem [{\citenamefont {Ziegler}\ and\ \citenamefont {Nichols}(1942)}]{Ziegler1942}%
  \BibitemOpen
  \bibfield  {author} {\bibinfo {author} {\bibfnamefont {J.~G.}\ \bibnamefont {Ziegler}}\ and\ \bibinfo {author} {\bibfnamefont {N.~B.}\ \bibnamefont {Nichols}},\ }\href@noop {} {\bibfield  {journal} {\bibinfo  {journal} {Transactions of the American society of mechanical engineers}\ }\textbf {\bibinfo {volume} {64}},\ \bibinfo {pages} {759} (\bibinfo {year} {1942})}\BibitemShut {NoStop}%
\bibitem [{\citenamefont {Sallab}\ \emph {et~al.}(2017)\citenamefont {Sallab} \emph {et~al.}}]{Driving2017}%
  \BibitemOpen
  \bibfield  {author} {\bibinfo {author} {\bibfnamefont {A.~E.}\ \bibnamefont {Sallab}} \emph {et~al.},\ }\href@noop {} {\bibfield  {journal} {\bibinfo  {journal} {Electronic Imaging}\ }\textbf {\bibinfo {volume} {29}},\ \bibinfo {pages} {70–76} (\bibinfo {year} {2017})}\BibitemShut {NoStop}%
\bibitem [{\citenamefont {Kiran}\ \emph {et~al.}(2021)\citenamefont {Kiran} \emph {et~al.}}]{Driving2021}%
  \BibitemOpen
  \bibfield  {author} {\bibinfo {author} {\bibfnamefont {B.~R.}\ \bibnamefont {Kiran}} \emph {et~al.},\ }\href@noop {} {\enquote {\bibinfo {title} {Deep reinforcement learning for autonomous driving: A survey},}\ } (\bibinfo {year} {2021}),\ \Eprint {http://arxiv.org/abs/2002.00444} {arXiv:2002.00444 [cs.LG]} \BibitemShut {NoStop}%
\bibitem [{\citenamefont {Wei}, \citenamefont {Wang},\ and\ \citenamefont {Zhu}(2017)}]{Wei}%
  \BibitemOpen
  \bibfield  {author} {\bibinfo {author} {\bibfnamefont {T.}~\bibnamefont {Wei}}, \bibinfo {author} {\bibfnamefont {Y.}~\bibnamefont {Wang}}, \ and\ \bibinfo {author} {\bibfnamefont {Q.}~\bibnamefont {Zhu}},\ }in\ \href@noop {} {\emph {\bibinfo {booktitle} {Proceedings of the 54th Annual Design Automation Conference 2017}}},\ \bibinfo {series and number} {DAC '17}\ (\bibinfo  {publisher} {Association for Computing Machinery},\ \bibinfo {address} {New York, NY, USA},\ \bibinfo {year} {2017})\BibitemShut {NoStop}%
\bibitem [{\citenamefont {Mocanu}\ \emph {et~al.}(2019)\citenamefont {Mocanu} \emph {et~al.}}]{Energy}%
  \BibitemOpen
  \bibfield  {author} {\bibinfo {author} {\bibfnamefont {E.}~\bibnamefont {Mocanu}} \emph {et~al.},\ }\href@noop {} {\bibfield  {journal} {\bibinfo  {journal} {IEEE Transactions on Smart Grid}\ }\textbf {\bibinfo {volume} {10}},\ \bibinfo {pages} {3698} (\bibinfo {year} {2019})}\BibitemShut {NoStop}%
\bibitem [{\citenamefont {Bøhn}\ \emph {et~al.}(2019)\citenamefont {Bøhn}, \citenamefont {Coates}, \citenamefont {Moe},\ and\ \citenamefont {Johansen}}]{Bohn2019}%
  \BibitemOpen
  \bibfield  {author} {\bibinfo {author} {\bibfnamefont {E.}~\bibnamefont {Bøhn}}, \bibinfo {author} {\bibfnamefont {E.~M.}\ \bibnamefont {Coates}}, \bibinfo {author} {\bibfnamefont {S.}~\bibnamefont {Moe}}, \ and\ \bibinfo {author} {\bibfnamefont {T.~A.}\ \bibnamefont {Johansen}},\ }\href {\doibase 10.1109/ICUAS.2019.8798254} {\enquote {\bibinfo {title} {Deep reinforcement learning attitude control of fixed-wing uavs using proximal policy optimization},}\ } (\bibinfo {year} {2019})\BibitemShut {NoStop}%
\bibitem [{\citenamefont {Qin}\ \emph {et~al.}(2018)\citenamefont {Qin}, \citenamefont {Zhang}, \citenamefont {Shi},\ and\ \citenamefont {Jinglong}}]{DRPID}%
  \BibitemOpen
  \bibfield  {author} {\bibinfo {author} {\bibfnamefont {Y.}~\bibnamefont {Qin}}, \bibinfo {author} {\bibfnamefont {W.}~\bibnamefont {Zhang}}, \bibinfo {author} {\bibfnamefont {J.}~\bibnamefont {Shi}}, \ and\ \bibinfo {author} {\bibfnamefont {L.}~\bibnamefont {Jinglong}}\ }(\bibinfo {year} {2018})\ pp.\ \bibinfo {pages} {1--6}\BibitemShut {NoStop}%
\bibitem [{\citenamefont {Ming}\ \emph {et~al.}(2023)\citenamefont {Ming}, \citenamefont {Liu}, \citenamefont {Li}, \citenamefont {Yin},\ and\ \citenamefont {Zhang}}]{SACaircraft2023}%
  \BibitemOpen
  \bibfield  {author} {\bibinfo {author} {\bibfnamefont {R.}~\bibnamefont {Ming}}, \bibinfo {author} {\bibfnamefont {X.}~\bibnamefont {Liu}}, \bibinfo {author} {\bibfnamefont {Y.}~\bibnamefont {Li}}, \bibinfo {author} {\bibfnamefont {Y.}~\bibnamefont {Yin}}, \ and\ \bibinfo {author} {\bibfnamefont {W.}~\bibnamefont {Zhang}},\ }\href {\doibase 10.1007/s10489-023-04876-y} {\bibfield  {journal} {\bibinfo  {journal} {Applied Intelligence}\ } (\bibinfo {year} {2023}),\ 10.1007/s10489-023-04876-y}\BibitemShut {NoStop}%
\bibitem [{\citenamefont {Schulman}\ \emph {et~al.}(2017{\natexlab{a}})\citenamefont {Schulman}, \citenamefont {Wolski}, \citenamefont {Dhariwal}, \citenamefont {Radford},\ and\ \citenamefont {Klimov}}]{PPO}%
  \BibitemOpen
  \bibfield  {author} {\bibinfo {author} {\bibfnamefont {J.}~\bibnamefont {Schulman}}, \bibinfo {author} {\bibfnamefont {F.}~\bibnamefont {Wolski}}, \bibinfo {author} {\bibfnamefont {P.}~\bibnamefont {Dhariwal}}, \bibinfo {author} {\bibfnamefont {A.}~\bibnamefont {Radford}}, \ and\ \bibinfo {author} {\bibfnamefont {O.}~\bibnamefont {Klimov}},\ }\href@noop {} {\enquote {\bibinfo {title} {Proximal policy optimization algorithms},}\ } (\bibinfo {year} {2017}{\natexlab{a}}),\ \Eprint {http://arxiv.org/abs/1707.06347} {arXiv:1707.06347 [cs.LG]} \BibitemShut {NoStop}%
\bibitem [{\citenamefont {Lakhani}, \citenamefont {Chowdhury},\ and\ \citenamefont {Lu}(2021)}]{Self-preserving}%
  \BibitemOpen
  \bibfield  {author} {\bibinfo {author} {\bibfnamefont {A.~I.}\ \bibnamefont {Lakhani}}, \bibinfo {author} {\bibfnamefont {M.~A.}\ \bibnamefont {Chowdhury}}, \ and\ \bibinfo {author} {\bibfnamefont {Q.}~\bibnamefont {Lu}},\ }\href {\doibase 10.48550/ARXIV.2112.15187} {\enquote {\bibinfo {title} {Stability-preserving automatic tuning of pid control with reinforcement learning},}\ } (\bibinfo {year} {2021})\BibitemShut {NoStop}%
\bibitem [{\citenamefont {Carlucho}, \citenamefont {{De Paula}},\ and\ \citenamefont {Acosta}(2020)}]{MIMOPID}%
  \BibitemOpen
  \bibfield  {author} {\bibinfo {author} {\bibfnamefont {I.}~\bibnamefont {Carlucho}}, \bibinfo {author} {\bibfnamefont {M.}~\bibnamefont {{De Paula}}}, \ and\ \bibinfo {author} {\bibfnamefont {G.~G.}\ \bibnamefont {Acosta}},\ }\href@noop {} {\bibfield  {journal} {\bibinfo  {journal} {ISA Transactions}\ }\textbf {\bibinfo {volume} {102}},\ \bibinfo {pages} {280} (\bibinfo {year} {2020})}\BibitemShut {NoStop}%
\bibitem [{\citenamefont {Silver}\ \emph {et~al.}(2014)\citenamefont {Silver} \emph {et~al.}}]{DDPG1}%
  \BibitemOpen
  \bibfield  {author} {\bibinfo {author} {\bibnamefont {Silver}} \emph {et~al.},\ }\href@noop {} {\bibfield  {journal} {\bibinfo  {journal} {31st International Conference on Machine Learning, ICML 2014}\ }\textbf {\bibinfo {volume} {1}} (\bibinfo {year} {2014})}\BibitemShut {NoStop}%
\bibitem [{\citenamefont {Lillicrap}\ \emph {et~al.}(2019)\citenamefont {Lillicrap} \emph {et~al.}}]{DDPG2}%
  \BibitemOpen
  \bibfield  {author} {\bibinfo {author} {\bibfnamefont {T.~P.}\ \bibnamefont {Lillicrap}} \emph {et~al.},\ }\href@noop {} {\enquote {\bibinfo {title} {Continuous control with deep reinforcement learning},}\ } (\bibinfo {year} {2019}),\ \Eprint {http://arxiv.org/abs/1509.02971} {arXiv:1509.02971 [cs.LG]} \BibitemShut {NoStop}%
\bibitem [{\citenamefont {Sutton}, \citenamefont {Precup},\ and\ \citenamefont {Singh}(1999)}]{Sutton1999Semi-Markov}%
  \BibitemOpen
  \bibfield  {author} {\bibinfo {author} {\bibfnamefont {R.~S.}\ \bibnamefont {Sutton}}, \bibinfo {author} {\bibfnamefont {D.}~\bibnamefont {Precup}}, \ and\ \bibinfo {author} {\bibfnamefont {S.}~\bibnamefont {Singh}},\ }\href@noop {} {\bibfield  {journal} {\bibinfo  {journal} {Artificial intelligence}\ }\textbf {\bibinfo {volume} {112}},\ \bibinfo {pages} {181} (\bibinfo {year} {1999})}\BibitemShut {NoStop}%
\bibitem [{\citenamefont {Schulman}\ \emph {et~al.}(2017{\natexlab{b}})\citenamefont {Schulman} \emph {et~al.}}]{TRPO}%
  \BibitemOpen
  \bibfield  {author} {\bibinfo {author} {\bibfnamefont {J.}~\bibnamefont {Schulman}} \emph {et~al.},\ }\href@noop {} {\enquote {\bibinfo {title} {Trust region policy optimization},}\ } (\bibinfo {year} {2017}{\natexlab{b}}),\ \Eprint {http://arxiv.org/abs/1502.05477} {arXiv:1502.05477 [cs.LG]} \BibitemShut {NoStop}%
\bibitem [{\citenamefont {Haarnoja}\ \emph {et~al.}(2018)\citenamefont {Haarnoja}, \citenamefont {Zhou}, \citenamefont {Abbeel},\ and\ \citenamefont {Levine}}]{SAC}%
  \BibitemOpen
  \bibfield  {author} {\bibinfo {author} {\bibfnamefont {T.}~\bibnamefont {Haarnoja}}, \bibinfo {author} {\bibfnamefont {A.}~\bibnamefont {Zhou}}, \bibinfo {author} {\bibfnamefont {P.}~\bibnamefont {Abbeel}}, \ and\ \bibinfo {author} {\bibfnamefont {S.}~\bibnamefont {Levine}},\ }in\ \href {https://proceedings.mlr.press/v80/haarnoja18b.html} {\emph {\bibinfo {booktitle} {Proceedings of the 35th International Conference on Machine Learning}}},\ \bibinfo {series} {Proceedings of Machine Learning Research}, Vol.~\bibinfo {volume} {80},\ \bibinfo {editor} {edited by\ \bibinfo {editor} {\bibfnamefont {J.}~\bibnamefont {Dy}}\ and\ \bibinfo {editor} {\bibfnamefont {A.}~\bibnamefont {Krause}}}\ (\bibinfo  {publisher} {PMLR},\ \bibinfo {year} {2018})\ pp.\ \bibinfo {pages} {1861--1870}\BibitemShut {NoStop}%
\bibitem [{\citenamefont {Nian}, \citenamefont {Liu},\ and\ \citenamefont {Huang}(2020)}]{NIAN2020}%
  \BibitemOpen
  \bibfield  {author} {\bibinfo {author} {\bibfnamefont {R.}~\bibnamefont {Nian}}, \bibinfo {author} {\bibfnamefont {J.}~\bibnamefont {Liu}}, \ and\ \bibinfo {author} {\bibfnamefont {B.}~\bibnamefont {Huang}},\ }\href {\doibase https://doi.org/10.1016/j.compchemeng.2020.106886} {\bibfield  {journal} {\bibinfo  {journal} {Computers \& Chemical Engineering}\ }\textbf {\bibinfo {volume} {139}},\ \bibinfo {pages} {106886} (\bibinfo {year} {2020})}\BibitemShut {NoStop}%
\bibitem [{\citenamefont {Henderson}\ \emph {et~al.}(2018)\citenamefont {Henderson}, \citenamefont {Islam}, \citenamefont {Bachman}, \citenamefont {Pineau}, \citenamefont {Precup},\ and\ \citenamefont {Meger}}]{Henderson2018}%
  \BibitemOpen
  \bibfield  {author} {\bibinfo {author} {\bibfnamefont {P.}~\bibnamefont {Henderson}}, \bibinfo {author} {\bibfnamefont {R.}~\bibnamefont {Islam}}, \bibinfo {author} {\bibfnamefont {P.}~\bibnamefont {Bachman}}, \bibinfo {author} {\bibfnamefont {J.}~\bibnamefont {Pineau}}, \bibinfo {author} {\bibfnamefont {D.}~\bibnamefont {Precup}}, \ and\ \bibinfo {author} {\bibfnamefont {D.}~\bibnamefont {Meger}}\ }(\bibinfo  {publisher} {AAAI Press},\ \bibinfo {year} {2018})\BibitemShut {NoStop}%
\bibitem [{\citenamefont {Mnih}\ \emph {et~al.}(2015)\citenamefont {Mnih} \emph {et~al.}}]{mnih2015humanlevel}%
  \BibitemOpen
  \bibfield  {author} {\bibinfo {author} {\bibfnamefont {V.}~\bibnamefont {Mnih}} \emph {et~al.},\ }\href@noop {} {\bibfield  {journal} {\bibinfo  {journal} {Nature}\ }\textbf {\bibinfo {volume} {518}},\ \bibinfo {pages} {529} (\bibinfo {year} {2015})}\BibitemShut {NoStop}%
\bibitem [{\citenamefont {Kingma}\ and\ \citenamefont {Ba}(2017)}]{kingma2017adam}%
  \BibitemOpen
  \bibfield  {author} {\bibinfo {author} {\bibfnamefont {D.~P.}\ \bibnamefont {Kingma}}\ and\ \bibinfo {author} {\bibfnamefont {J.}~\bibnamefont {Ba}},\ }\href@noop {} {\enquote {\bibinfo {title} {Adam: A method for stochastic optimization},}\ } (\bibinfo {year} {2017})\BibitemShut {NoStop}%
\end{thebibliography}
\end{document}